\documentclass[conference]{IEEEtran}
\IEEEoverridecommandlockouts
\usepackage{cite}
\usepackage{amsmath,amssymb,amsfonts}
\usepackage{graphicx}
\usepackage{textcomp}
\usepackage[table,xcdraw]{xcolor}
\usepackage{xcolor}
\usepackage[nameinlink,capitalise]{cleveref}
\usepackage[algoruled,boxed,lined]{algorithm2e}
\usepackage{graphicx}
\usepackage{amsmath}
\usepackage{multirow}
\usepackage{amsthm} 
\usepackage{lipsum}
\usepackage{flushend}
\usepackage[export]{adjustbox}

\usepackage{url}
\def\BibTeX{{\rm B\kern-.05em{\sc i\kern-.025em b}\kern-.08em
    T\kern-.1667em\lower.7ex\hbox{E}\kern-.125emX}}

\begin{document}

\title{A Comparative Study of  U-Net Topologies for Background Removal in Histopathology Images }

\author{\IEEEauthorblockN{Abtin Riasatian,  Maral Rasoolijaberi,  Morteza Babaei,  H.R. Tizhoosh}
\thanks{* This work was funded by a  NSERC-CRD grant on ``Design and Development of Devices and Procedures for Recognizing Artefacts and Foreign
Tissue Origin for Diagnostic Pathology''}
\IEEEauthorblockA{\textit{Kimia Lab, University of Waterloo, Canada} \\
\{abtin.riasatian, mrasooli, mbabaie, tizhoosh\}@uwaterloo.ca}

}

\maketitle

\begin{abstract}
During the last decade, the digitization of pathology has gained considerable momentum. Digital pathology offers many advantages including more efficient workflows, easier collaboration as well as a powerful venue for telepathology. At the
same time, applying Computer-Aided Diagnosis (CAD) on Whole
Slide Images (WSIs) has received substantial attention as a direct result of the digitization. The first step in any image analysis is to extract the tissue. Hence, background removal is an essential prerequisite for efficient
and accurate results for many algorithms. In spite of the obvious discrimination for human operator,  the identification of tissue regions in WSIs could be challenging for computers 
mainly due to the existence of color variations and artifacts. Moreover, some cases such as alveolar tissue
types, fatty tissues, and tissues with poor staining are difficult to
detect. In this paper, we perform experiments on U-Net architecture with different network backbones (different topologies) to remove the background as well as artifacts from WSIs in order to extract the tissue regions. We compare a wide range of backbone networks including MobileNet, VGG16, EfficientNet-B3,
ResNet50, ResNext101 and DenseNet121. We trained and evaluated the network on a manually labeled
subset of The Cancer Genome Atlas (TCGA) Dataset. EfficientNet-B3 and MobileNet by almost $99\%$ sensitivity and specificity reached the best results. 
\end{abstract}


\section{Introduction}
In the recent decade, the image digitization has recently become more popular in the pathology practice. Improvement in
this technology has led to the manufacturing of high-resolution
whole-slide scanners which can produce WSIs in a short time.
The digital scan of the biopsy glass slides can be explored by image
viewers rather than the conventional microscope. Also, despite the large size of scans (a typical WSI file is usually at least several hundred megabytes), new storage
and network sharing progress make it possible to share these
files much faster than mailing glass samples for the
purpose of consultations and acquiring second opinions \cite{al2012digital}. An important benefit of digital pathology is that AI and computer vision methods can be applied on tissue scans to help pathologists create more accurate reports \cite{pantanowitz2011review}. Due  to  the  large size of WSIs, most pathology image processing methods divide the slides into small tiles (patches) before feeding them to the CAD systems. Unquestionably foreground segmentation is a necessary prerequisite for almost every tile-based method to decrease the time complexity and possibility of making mistakes by the algorithms due to analyzing irrelevant parts \cite{ kumar2018deep}. Thus, one has to remove irrelevant pixels from WSIs as much without removing any tissue pixels  \cite{rajalocalization,khan2012ranpec}. Since in medical imaging, histopathology image analysis is generally the last step for cancer diagnosis \cite{he2012histology}, it is crucial to avoid losing tissue pixels. Therefore, the expected segmentation sensitivity has to be very high.  




Another application of tissue foreground segmentation is in whole slide scanners which digitize glass slides containing tissue specimens to generate WSI files. The focus depth of whole slide scanners must be adjusted for different tissue regions due to variable tissue thickness. Hence, scanners need to identify all areas which contain tissue. If an error occurs during digitizing glass slides, there is no way to fix the error in the following  steps of the digital pathology workflow. Currently, a technician manually checks every slide after scanning, which is a tedious and expensive procedure \cite{pantanowitz2011review}, \cite{bandi2017comparison}.

\begin{figure}[t]
\centerline{\includegraphics[width=\columnwidth]{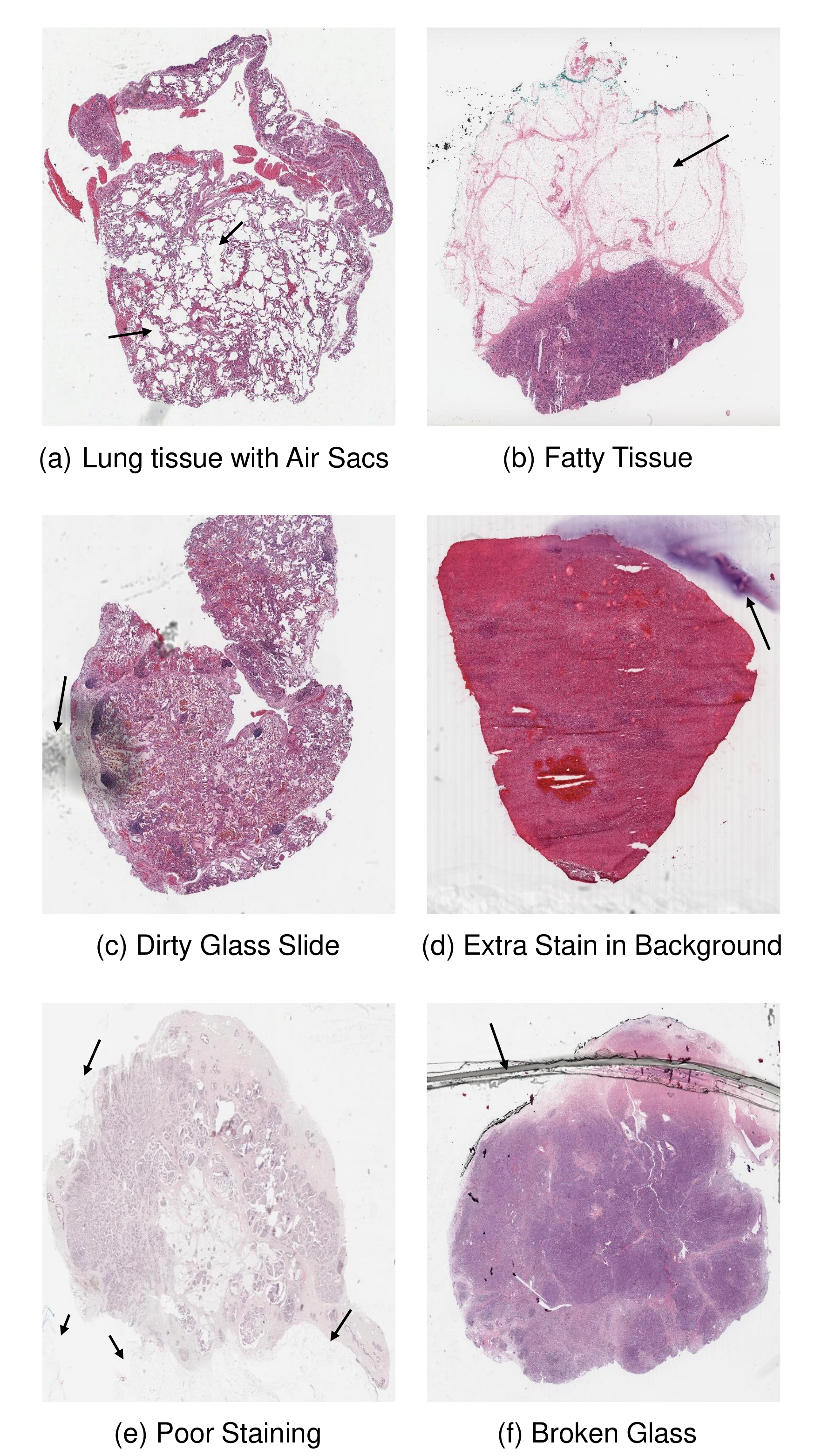}}
\caption{Some examples for challenges in tissue extraction (images selected from the TCGA dataset).}
\label{fig:challing_cases}
\end{figure}

Some of the challenges in tissue segmentation in histopathology images are related to the tissue type. For instance, air sacs in  the lung, and fat which could appear in many tissue types, may confuse algorithms due to their resemblance with the background color while they can be easily segmented as tissue by an expert. Another important challenge is the presence of artifacts including bubbles, tissue folds, extra stain, broken glass, debris, and marker traces \cite{rajalocalization, babaie2019deep}. Moreover, mistakes in tissue preparation such as weak staining raise difficulties for tissue identification algorithms  \cite{pantanowitz2011review, kothari2013eliminating}. 
Examples of some of the mentioned challenging cases are indicated in Fig.\ref{fig:challing_cases}.

In this paper, we propose a novel method to identify tissue areas in the WSI thumbnail images. The main contributions of this paper are:
(1) Releasing a publicly available dataset consisting of 244 thumbnails of TCGA WSIs along with their segmentation masks,
(2) proposing a deep learning topology using  U-Net for reliable, accurate, and automatic tissue segmentation, and (3) comparing the performance of different encoders as the backbone of U-Net in the tissue segmentation task. 
The manifest to download the data from the GDC website, manually refined labels ,as well as codes to run the proposed U-Net, is available for download\footnote{https://kimialab.uwaterloo.ca/kimia/index.php/data-and-code/}.

\section{Related literature}
The identification of regions containing tissue is usually the first step in histopathology image analysis. However, this problem is often treated  as a trivial part of research mostly solved via threshold-based methods.
Most research papers have used empirical  rules to set the threshold for different image specifications such as gradient, intensity, color, etc.  \cite{oswal2013entropy,Babaie_2017_CVPR_Workshops,bentaieb2018predicting, kothari2013eliminating, bug2015foreground}. 
\subsection{Machine vision based methods}

Estimation of the texture complexity in small neighbors has been used by Oswal et al. to detect the foreground \cite{oswal2013entropy}. Babaie et al. \cite{Babaie_2017_CVPR_Workshops} used homogeneity and gradient values to estimate the patch complexity. Bentaieb et al. \cite{bentaieb2018predicting} used a threshold on the pixel intensity values to detect the tissue. As another example, Kothari et al. \cite{kothari2013eliminating} removed blank regions by setting a threshold on saturation and intensity of pixels.  
Other works used homogeneity criteria to only select  patches containing a considerable part of the tissue \cite{erfankhah2019heterogeneity}.

The Otsu algorithm \cite{otsu1982role} as a robust iterative thresholding method has been widely used to compute the optimal threshold. Mohit employed the Otsu method on HSV transformed image for background removal  \cite{mohit2017automated}. Nguyen et al. applied Otsu's method on the b channel of the LAB color space to obtain tissue regions in WSIs \cite{nguyen2011prostate}.

One of the most well known and widely used open-source libraries in digital histopathology,  Histomics Toolkit (HistomicsTK)\footnote{https://github.com/DigitalSlideArchive/HistomicsTK} can also perform tissue detection on the thumbnail of a WSI. The process contains a series of Gaussian smoothing and Otsu thresholding. Also, another threshold is used to filter regions smaller than a preset size.

In contrast to the mentioned works, which treat background detection as a small part of the entire WSI processing, there are few studies which 
 have addressed the foreground/background detection  in histopathology slides as a major  problem \cite{rajalocalization, bug2015foreground, bandi2017comparison, fouad2017unsupervised}. Similar to the previous vision-based methods, FESI \cite{bug2015foreground} used a combination of basic methods, such as median filtering, thresholding, erosion and dilation to address this problem. Calculation of absolute value of the Laplacian based on gray-scale image, and then applying Gaussian filter  is used in their work. Recently Chen et al. \cite{chen2019tissueloc} introduced tissue localization method by applying inverse binarization on the gray-scale images followed by erosion and dilation.

\subsection{Network based Methods}
\begin{figure}[ht]
\centerline{\includegraphics[width=\columnwidth]{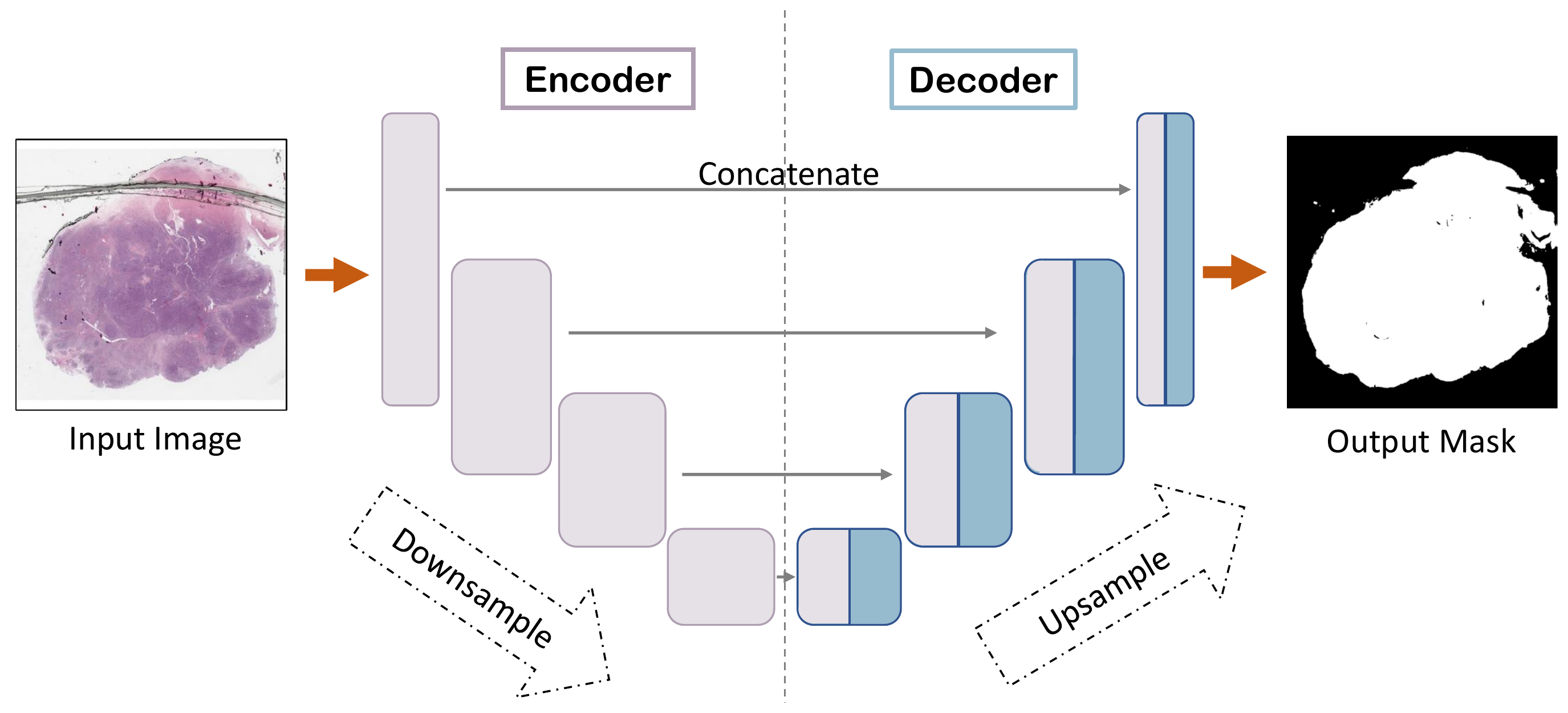}} 
\caption{Network Architecture: Each block shows a feature map. }
\label{fig:U-Net}
\end{figure}

Neural network based methods are a rather recent trend in the literature to address the tissue segmentation. Raja et al. \cite{rajalocalization} have extracted four different features including color, appearance, texture and spatial features. They fed the selected features to a two-layer neural network to classify the patches into background and foreground pixels. Bandi et al. \cite{bandi2017comparison} trained FCN and U-Net networks for tissue segmentation with patches with a single label. They used patches with the size of $892 \times 892$ pixels for U-Net and $128 \times 128$ pixels for FCNN. Their patches were randomly extracted from 54 WSIs.  They assigned only one label to each patch based on its central pixel which means the same labels are allocated to roughly 800,000 pixels in the U-Net case. It seems that all  network-based methods are working on the highest usually available magnification (namely, $20 \times$ magnification). As a result, for a whole slide processing, a large number of small patches must be fed to their network which is a time-consuming task. 
However, a more efficient way of segmentation is to assign a label to each pixel in a thumbnail to save time and also to avoid losing tissue parts (especially borders)  as much as possible. Therefore, in this paper we provide manually labeled WSI thumbnails (low magnification) to train U-Net models (Fig. \ref{fig:U-Net}). We have compared the most commonly used network architectures to find the best backbone for proposed U-Net.

\subsection{U-Net}
U-Net is a convolutional neural network which firstly was  proposed for the segmentation of neural structures in electron microscopic images in 2015 \cite{ronneberger2015u}.  Since then, this network has shown impressive performance in various segmentation tasks in medical imaging. Dong et al. \cite{dong2017automatic} proposed an automatic method to detect and segment brain tumors in MRI by using U-Net. Bulten et al. \cite{bulten2018automated} utilized U-Net for epithelial tissue segmentation to assist pathologists in prostate cancer diagnosis. Naylor et al.  \cite{naylor2018segmentation} proposed a method for cell nuclei segmentation by formulating this task as the regression of the distance map. They compared results of three different architectures: (1) the pre-trained VGG16 with fine-tuning as the FCN approach, (2) U-Net, and (3) Mask R-CNN with the pre-trained ResNet 101 as its backbone. U-Net can be trained end-to-end using a small number of images \cite{Yakubovskiy:2019}. This is the most significant  advantage  of the U-Net, especially in applications such as biomedical domain where usually only a few annotated images are available.


\section{Methodology}

\subsection{Data Annotation}
For tissue segmentation, a label must be assigned to each pixel to indicate whether it belongs to a tissue region or not. A mask is a binary image where every pixel is either zero (black) or one (white) where the white pixels generally mark the region of interest. 


A typical WSI contains more than several thousand pixels in each image axis (i.e., a typical WSI may easily be $50,000 \times 50,000$ or larger). Thus, assigning a label to each pixel of the WSI is not a feasible task. To overcome this challenge, we first work with thumbnails instead of WSIs, that is generally the image at $1\times$ magnification. Working with thumbnails has the advantage of fast computation. Also, tissue regions at higher magnifications can be constructed from their corresponding segmented thumbnails by simple calculations commonly known for the pyramidal structures of whole slide images. 
We developed a handcrafted image processing approach, details in Alg. \ref{Algorithm1}, to produce initial masks from thumbnails  \cite{grayscal_regions}, \cite{belsare2012histopathological}.

\begin{algorithm}[t]
\SetAlgoLined

\SetKwData{binThmb}{binThmb}
\SetKwData{rgbThmb}{rgbThmb}
\SetKwData{contours}{contours}
\SetKwData{hierarchy}{hierarchy}
\SetKwData{chosenContours}{chosenContours}
\SetKwData{fatherContours}{fatherContours}
\SetKwData{firstLevelChildren}{firstLevelChildren}
\SetKwData{areaThreshold}{areaThreshold}
\SetKwData{ratioThreshold}{ratioThreshold}
\SetKwData{distThreshold}{distThreshold}
\SetKwData{distCond}{distCond}
\SetKwData{areaCond}{areaCond}
\SetKwData{finalMask}{finalMask}
\SetKwData{hole}{hole}
\SetKwData{hUpperThresh}{hUpperThresh}
\SetKwData{hLowerThresh}{hLowerThresh}
\SetKwData{i}{i}
\SetKwData{x}{x}

\SetKwFunction{binaryThresholding}{binaryThresholding}
\SetKwFunction{findContours}{findContours}
\SetKwFunction{getContours}{getContours}
\SetKwFunction{append}{append}
\SetKwFunction{min}{min}
\SetKwFunction{sort}{sort}
\SetKwFunction{invert}{invert}
\SetKwFunction{drawContours}{drawContours}
\SetKwFunction{distanceToClosest}{distanceToClosest}
\SetKwFunction{getArea}{getArea}

\SetKwInOut{Input}{Input}
\SetKwInOut{Output}{Output}
\Input{The rgb thumbnail of the WSI}
\Output{Thumbnail binary mask with the same size}
\BlankLine
\chosenContours $\leftarrow$ []\;
\binThmb $\leftarrow$ \binaryThresholding{\rgbThmb}\;
\contours, \hierarchy $\leftarrow$ \findContours{\binThmb}\;
\BlankLine
\fatherContours $\leftarrow$ \getContours{\contours, \hierarchy, $0$}\;
\BlankLine
\append{\chosenContours, \fatherContours}\;
\BlankLine
\firstLevelChildren $\leftarrow$ \getContours{\contours, \hierarchy, $1$}\;
\BlankLine
\firstLevelChildren $\leftarrow$ \sort{\firstLevelChildren, 'area'}\;
\append{\chosenContours, \firstLevelChildren$[0]$}\;
\BlankLine
i $\leftarrow$ 1\;
\While{\firstLevelChildren$[\i].area$ \textgreater \min{\firstLevelChildren$[\i-1].area$ * \ratioThreshold, \areaThreshold}}{
    \append{\chosenContours, \firstLevelChildren$[i]$}\;
    \i $\leftarrow \i+1$
}
\BlankLine
\ForEach{\x in \firstLevelChildren}{
\BlankLine
\distCond $\leftarrow$ \distanceToClosest{\x, \chosenContours} \textless \distThreshold\;
\BlankLine
    \uIf{\distCond and \x not in \chosenContours}{
        \append{\chosenContours, \x}\;
    }
}
\BlankLine

\ForEach{\x in \firstLevelChildren}{
\BlankLine
\areaCond $\leftarrow$ \getArea{\x} \textgreater \areaThreshold\;
\BlankLine
    \uIf{\areaCond and \x not in \chosenContours}{
        \append{\chosenContours, \x}\;
    }
}

\drawContours{\chosenContours, \finalMask, 'white'}\;
\BlankLine

\BlankLine
\ForEach{\hole in \invert{\binThmb}}{
    \uIf{\hLowerThresh \textless \hole.area \textless \hUpperThresh}{
        \drawContours{\hole, \finalMask, 'black'}\;
    }
}

 \caption{Handcrafted Masking Method}
 \label{Algorithm1}
\end{algorithm}

\begin{figure*}[htb]
\centerline{\includegraphics[width=0.9\textwidth]{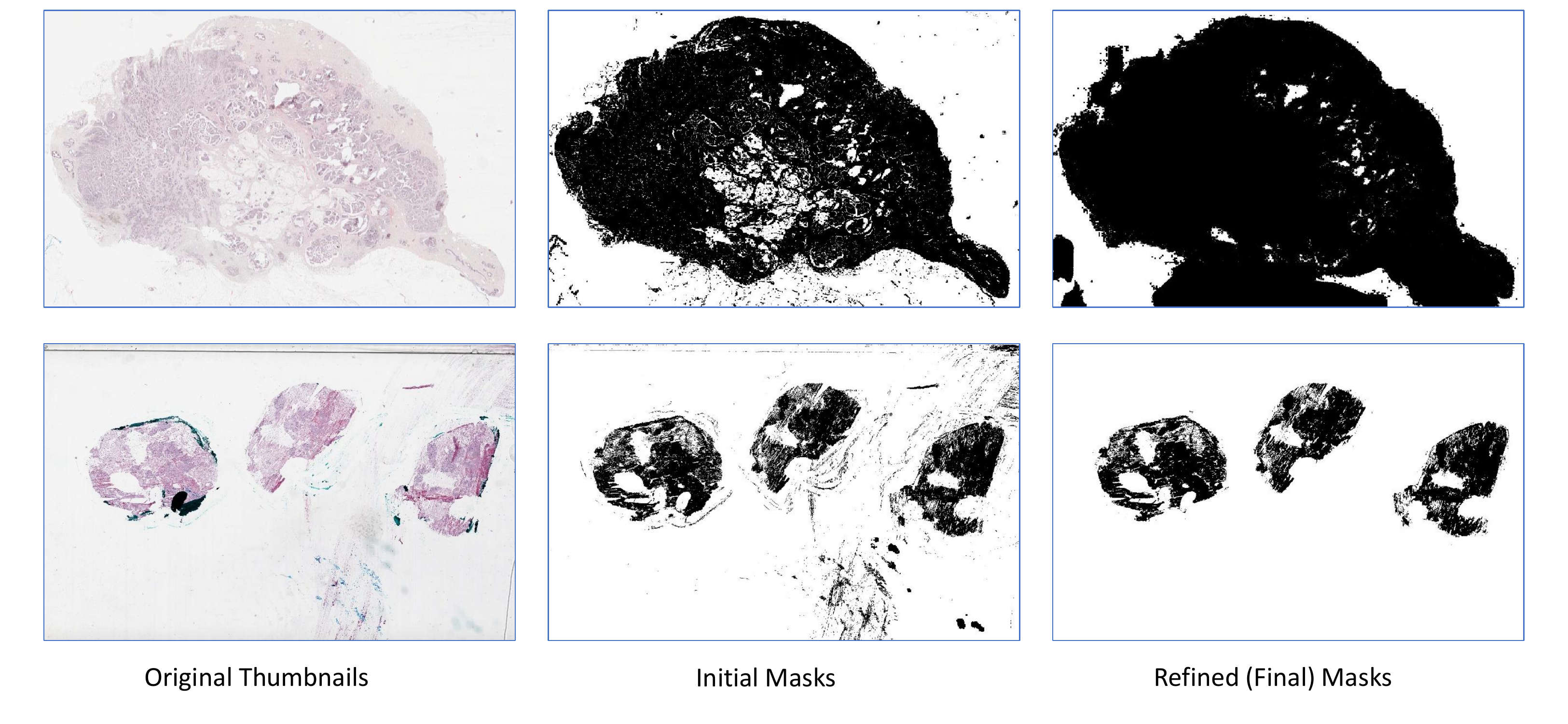}}
\caption{Two samples of sequential steps for generating masks.}
\label{fig:mask_refinement}
\end{figure*}

Masking was performed in $2.5 \times$ magnification to preserve details. Note that based on our practical experiments, in challenging cases, the $2.5 \times$ magnification was the smallest size which still could distinguish the tissue parts from artifacts such as extra staining. Thereafter, initial masks were refined manually to make sure that all tissue regions are selected, and noise and artifacts are removed as much as possible. An example of the mentioned steps can be found in Fig. \ref{fig:mask_refinement}.

With regard to difficult cases, we used image dilation with $3 \times 3$ kernels to make sure every pixel of tissue, especially at borders, are preserved. Finally, each pair of mask and thumbnail is resized (preserving the aspect ratio) in a way that each image dimension does not exceed $1024$ pixels to make the images  small enough to be processed by the network. It is noteworthy that the thumbnails have various dimensions necessitating background padding to have the unified size $1024 \times 1024$ for all images.

\subsection{Model Architecture} 
\label{model_Arch}
 U-Net, which is a fully convolutional network with a U-shape architecture, has two parts, called encoder and decoder. The first sub-network, known as the encoder, extracts high-level features to capture the image content. The decoder sub-network, also known as the expansion part, creates the desired segmentation map \cite{ronneberger2015u}. Fig. \ref{fig:U-Net} shows the proposed network architecture. U-Net-based deep networks, the same as U-Net, include two encoder and decoder sub-networks. As the input image passes through the first sub-network, higher-level features are extracted. In the next sub-network, deep  feature maps  are combined with low-level feature maps from the encoder sub-network. The spatial resolution of feature maps are increased in the second sub-network to achieve an output mask with the same size as the input image. The connections between the encoder and decoder in U-Net architecture facilitate information propagation. In terms of connections in the U-Net architecture, feature maps from the encoder part are cropped and concatenated to feature maps in the decoder sub-network to retrieve local information. These connections enable the network to learn from a few samples\cite{ronneberger2015u}.

To improve the performance of U-Net, we applied custom backbones on its architecture using Segmentation\_Models library\footnote{https://github.com/qubvel/segmentation\_models}. The encoder part of these customized networks are the feature extractor, i.e., complete network architecture except the last dense layer, of a chosen network, e.g., MobileNet. The decoder part consists of 5 decoder blocks with filters of size 256, 128, 64, 32 and 16 as it gets deeper. The structure of each decoder block is made up of one 2d-upsampling layer and two repetitions of 2d-convolution, batch-normalization and ReLU activation. Four skip connections connect layers from the encoder part, usually the output of ReLU activation at a certain layer of each encoder block, to the last four decoder blocks, after the up-sampling layer. The last layers of the network is a  2d-convolution layer with Sigmoid activation.

We experimented with six different backbones (topologies) for U-Net-based solutions for tissue segmentation which are introduced in Section \ref{section_experiment_training}.


\section{Experiments}

\subsection{Data}
We used 244 WSIs selected from different organs such as  
brain, breast, kidney, and lung. All WSIs were randomly selected from The Cancer Genoum Atlas (TCGA) dataset  \cite{web_TCGA_dataset, tomczak2015cancer}. TCGA is one of the largest publicly available datasets with histopathology whole slide images.

\subsection{Topologies and Training Process}\label{section_experiment_training}

We have experimented with various network topologies including MobileNet, VGG16, EfficientNetB3, ResNet50, ResNext101, and DenseNet121 as the backbone of U-Net model to find the most suitable ones for tissue segmentation  \cite{howard2017mobilenets, simonyan2014very, tan2019efficientnet, he2016deep, huang2017densely}. All  networks were trained for 50 epochs, no early stopping, using Adam optimizer \cite{kingma2014adam} with the learning rate of $1e-4$ on one NVIDIA Tesla V100 GPU with 32GB memory. After running the experiments with two loss functions, namely Jaccard Index and sensitivity plus specificity, we chose the latter so the network tries to come up with an approximation which avoids the misclassification of tissue parts as background while having a good performance at recognizing background. The drawback of using Jaccard Index as the loss function was the relatively low sensitivity of the results. The networks were initialized with ImageNet weights and trained and evaluated with five-fold cross validation. For each fold, 195 $1024 \times 1024$ RGB images were used as the input and binary masks with the same size as the label in which pixel value 1 (positive) meant tissue and pixel value 0 (negative) meant background. Input images and their corresponding masks were augmented by three transformations: (1) Random rotation within the range of -180 and 180 degrees, (2) random horizontal flipping, and (3) random vertical flipping. The  validation dataset contained 49 images for each fold.

\begin{table*}[h]
\centering
\caption{Summary of results: Comparing our networks with image-processing methods (gray rows).}
\begin{tabular}{l||c||c|c|c|c}
Method         & \begin{tabular}[c]{@{}c@{}}Time (s)\end{tabular} & \begin{tabular}[c]{@{}c@{}}Jaccard Index\end{tabular} & \begin{tabular}[c]{@{}c@{}}Dice  Coeff.\end{tabular} &
\begin{tabular}[c]{@{}c@{}}Sensitivity\end{tabular} & \begin{tabular}[c]{@{}c@{}}Specificity\end{tabular} \\ \hline
\rowcolor[HTML]{57EED4} 
MobileNet      & 0.11                                                & 0.95                                                     & 0.97                                                  & 0.99                                                    & 0.99                                                    \\
\rowcolor[HTML]{57EED4}
EfficientNet-B3 & 0.18                                                & 0.95                                                     & 0.97                                                  & 0.99                                                    & 0.98                                                    \\
ResNet50       & 0.16                                                & 0.94                                                     & 0.97                                                  & 0.99                                                    & 0.98                                                    \\

DenseNet121    & 0.16                                                & 0.93                                                     & 0.96                                                  & 0.99                                                    & 0.98                                                    \\

ResNext101     & 0.50                                                & 0.93                                                     & 0.96                                                  & 0.99                                                    & 0.98                                                    \\

VGG16          & 0.11                                                & 0.75                                                     & 0.82                                                  & 0.99                                                    & 0.81                                                    \\

\rowcolor[HTML]{D6DAD9} Improved FESI \cite{bug2015foreground}          & 0.11                                                & 0.86                                                     & 0.92                                                  & 0.91                                                    & 0.97                                                    \\

\rowcolor[HTML]{D6DAD9} TissueLoc \cite{chen2019tissueloc}          & 0.26                                                & 0.81                                                     & 0.88                                                  & 0.88                                                    & 0.97                                                    \\

\rowcolor[HTML]{D6DAD9} Otsu algorithm          & 0.02                                                & 0.81                                                     & 0.89                                                  & 0.82                                                    & 0.99                                                    \\
\rowcolor[HTML]{D6DAD9} Histomics-TK   & 0.13                                                   & 0.78                                                        & 0.87                                                    & 0.79                                                       & 0.99                                                       \\
\end{tabular}
\label{TABLE_RESULT}
\end{table*}


\subsection{Comparison of Methods}
To compare our results against other methods, we used the same input images fed to our networks as their input and calculated their performance against the ground-truth masks. All methods were checked to be able to work with the given inputs. We compared our results against four traditional computer vision methods:

(1) FESI algorithm \cite{bug2015foreground} is improved by changing the color space of the input image from BGR to LAB and the value of the first two channels, lightness and red/green value, are changed to maximum intensity value \footnote{ https://github.com/alexander-rakhlin/he\_stained\_fg\_extraction}. Color space of the resulting image is changed to gray-scale and binerized using the mean value of the image as threshold. This binary image is passed to the Gaussian filter instead of using the absolute value of the Laplacian of the gray-scale image as done in the original paper. 

(2) We used $locate\_tissue\_cnts$ function available in the open-source Python package\footnote{ https://github.com/PingjunChen/tissueloc}, TissueLoc\cite{chen2019tissueloc}, as a recently developed method for comparative purposes. We modified the function in a way that it uses the thumbnail image as input. Also all of  input parameters of the function are set to default values except $min\_tissue\_size$ which is set to 50 to make sure the algorithm would detect all tissue parts. 

(3) Histomics Toolkit Python library is one of the most popular libraries in the histopathology domain. \textit{saliency.tissue\_detection.get\_tissue\_mask} function was used as tissue segmentation method. We set the input parameters $deconvolve\_first$ to False, $n\_thresholding\_steps$ to 1 and $min\_size$ threshold to 50.

(4) Otsu binarization method as one of the well-known algorithms to classify pixels into foreground and background. The RGB thumbnail images are first converted to gray-scale and then the Otsu method is applied.

\subsection{Performance Evaluation}

In the test phase, we compared the result of all methods with the ground-truths via 5-fold cross validation. In addition to the processing time, four different performance measurements, i.e., Jaccard index, Dice coefficient, sensitivity, and specificity, were conducted which are defined as: 


\begin{align}
& \text{Jaccard} := \frac{\text{TP}}{\text{TP} + \text{FP} + \text{FN}}, \\
& \text{Dice} := \frac{2*\text{TP}}{2*\text{TP} + \text{FP} + \text{FN}}, \\
& \text{Sensitivity} := \frac{\text{TP}}{\text{TP} + \text{FN}}, \\
& \text{Specificity} := \frac{\text{TN}}{\text{TN} + \text{FP}},
\end{align}

\begin{figure}[b]
\centerline{\includegraphics[width=\columnwidth]{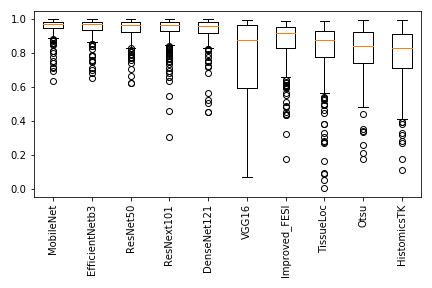}}
\caption{Jaccard Index}
\label{fig:jaccard_box_plot}
\end{figure}

where TP, TN, FP, and FN denote the number of true positives, true negatives, false positives, and false negatives, respectively. The segmented pixels are considered as positive and negative where they are labeled as tissue and non-tissue pixels, respectively. Note that the sensitivity is more important than the specificity in the tissue segmentation task because sensitivity penalizes wrong labeling of tissue region as background while specificity gives a penalty to the wrong labeling of the background region as tissue. In histopathology, it is paramount to do not miss any part of the tissue.

\begin{figure}[t]
\centerline{\includegraphics[width=\columnwidth]{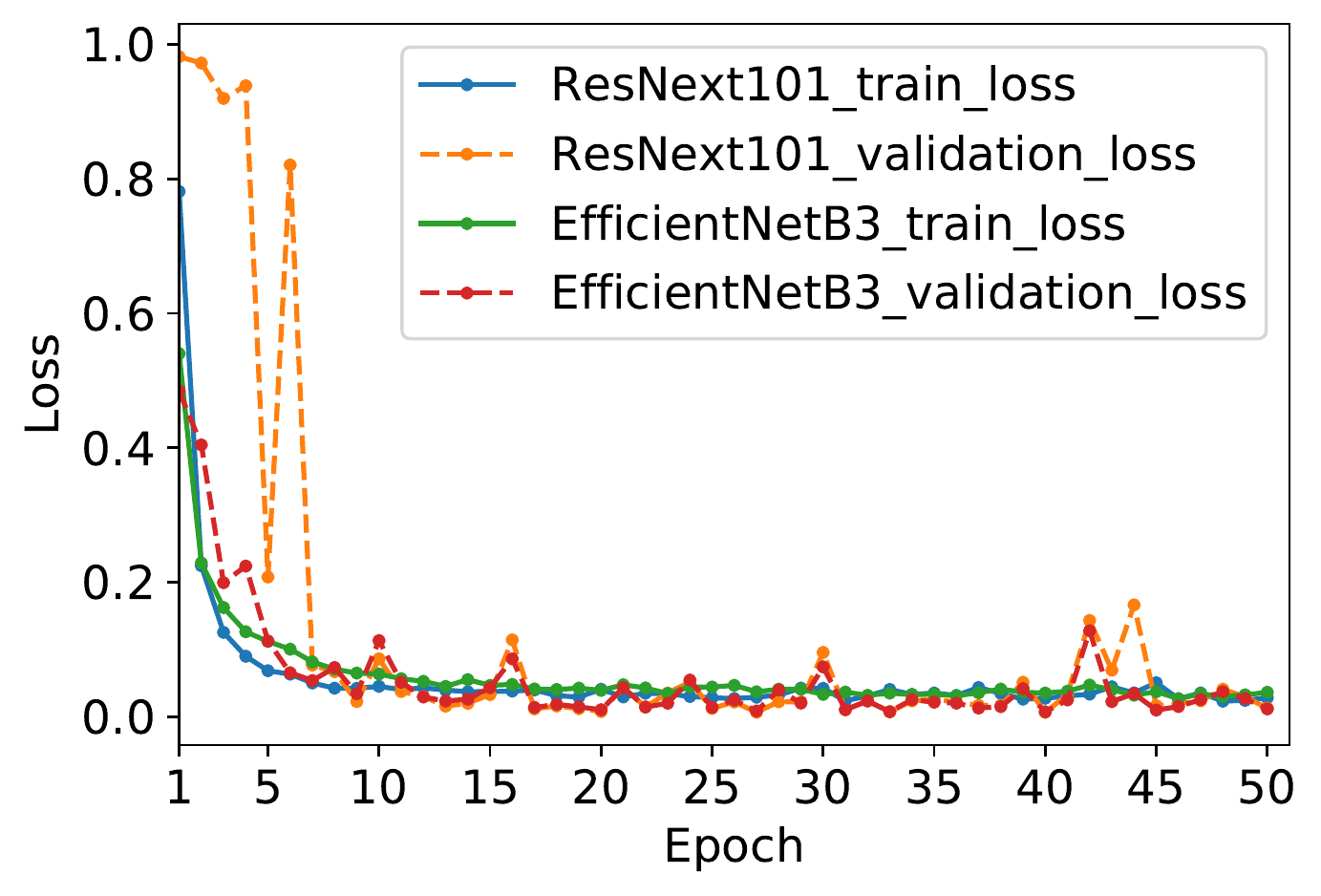}}
\caption{Losses for ResNext101 and EfficientNet-B3}
\label{fig:loss_plot}
\end{figure}

\subsection{Analysis of Results}
In addition to Improved FESI and TissueLoc methods, we chose HistomicsTK tissue segmentation and Otsu algorithm to compare our networks' results against well-known methods for histopathology image analysis. Table.\ref{TABLE_RESULT} shows that all networks, except VGG16, outperform all four handcrafted methods with respect to all performance measurements. The most important advantage of networks over the handcrafted methods is their high sensitivity ($\approx 99\%$). In addition networks are as fast as handcrafted methods such as Improved FESI and TissueLoc while achieving considerably higher Jaccard Index and Dice Coefficients. It can be seen that Jaccard Index for the networks with best performance, namely MobileNet and EfficientNet-B3, is 9\%  higher than the best handcrafted method, namely Improved FESI. Considering the changes in the validation loss for two networks, ResNext101 with around 51 million parameters and EfficientNet-B3 with less than 18 million parameters, through 50 epochs, Fig.\ref{fig:loss_plot}, it seems that both have the same pattern; 20 epochs appeared to be enough for proper  network training. This would take around 20 minutes for a medium-size network and 40 minutes for a large network which is a negligible cost considering the benefits of using networks.

\begin{figure*}[h]
\centerline{\includegraphics[width=0.9\textwidth]{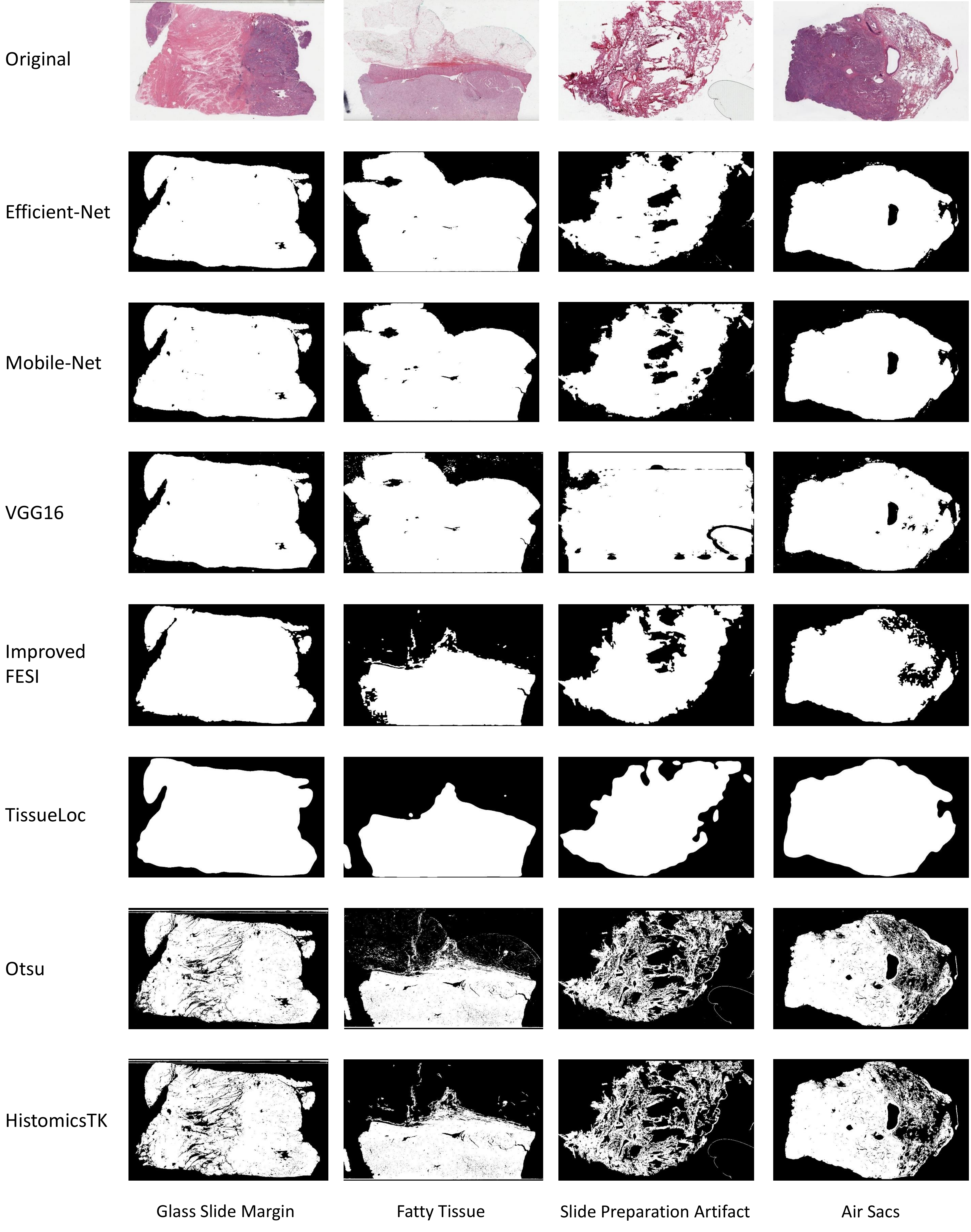}}
\caption{Example results of different methods in tissue segmentation task for challenging cases.}
\label{fig:resulting_masks}
\end{figure*}

To compare network backbones, it can be seen that MobileNet showed the best performance. Also EfficientNet-B3 shows very high performance. The poor performance of VGG16 could be due to several reasons. First of all, this network has a large number of parameters (more than 23 millions) while it only has 66 layers compared to other networks such as MobileNet with more than 8 million parameters and 128 layers and EfficientNet-B3 with around 18 million parameters and 418 layers. Also, the use of batch normalization and ReLU activation layers in the convolution blocks in other architectures have the benefits of avoiding internal covariate shift, which results in faster convergence, and keeping the network sparse, causing the generalization error to decrease, respectively. Since VGG16 lacks these layers in its architecture, it converges with difficulty.  Fig. \ref{fig:resulting_masks} depicts a visual overview of the proposed network results versus image-processing methods. As we can see, the proposed networks outperform in fatty tissue (second column) or tissue with an air bubble (third column) considerably. 



\section{Conclusion}
In this paper, we have compared the performance of U-Net with various custom topologies (backbones) for the identification of tissue regions in whole slide images. Using different networks combines the strength of current state-of-the-art CNNs with the custom architecture of the U-Net model for image segmentation. Whereas U-Net topologies can generate segments with 99\% sensitivity, handcrafted methods struggled to approach high 80\%. Both MobileNet and EfficientNet-B3 appeared to be the best backbone topology for the U-Net. 
The next step for this research would be changing the current binary masking network to a multi-class one which could label each pixel as classes such as marker trace, dirt and tissue fold, fat and informative tissue.
Authors have made the segmented images publicly available for sake of reproducibility.

\bibliographystyle{IEEEtran}

\bibliography{references}

\begin{thebibliography}{10}
\providecommand{\url}[1]{#1}
\csname url@samestyle\endcsname
\providecommand{\newblock}{\relax}
\providecommand{\bibinfo}[2]{#2}
\providecommand{\BIBentrySTDinterwordspacing}{\spaceskip=0pt\relax}
\providecommand{\BIBentryALTinterwordstretchfactor}{4}
\providecommand{\BIBentryALTinterwordspacing}{\spaceskip=\fontdimen2\font plus
\BIBentryALTinterwordstretchfactor\fontdimen3\font minus
  \fontdimen4\font\relax}
\providecommand{\BIBforeignlanguage}[2]{{%
\expandafter\ifx\csname l@#1\endcsname\relax
\typeout{** WARNING: IEEEtran.bst: No hyphenation pattern has been}%
\typeout{** loaded for the language `#1'. Using the pattern for}%
\typeout{** the default language instead.}%
\else
\language=\csname l@#1\endcsname
\fi
#2}}
\providecommand{\BIBdecl}{\relax}
\BIBdecl

\bibitem{al2012digital}
S.~Al-Janabi, A.~Huisman, and P.~J. Van~Diest, ``Digital pathology: current
  status and future perspectives,'' \emph{Histopathology}, vol.~61, no.~1, pp.
  1--9, 2012.

\bibitem{pantanowitz2011review}
L.~Pantanowitz, P.~N. Valenstein, A.~J. Evans, K.~J. Kaplan, J.~D. Pfeifer,
  D.~C. Wilbur, L.~C. Collins, and T.~J. Colgan, ``Review of the current state
  of whole slide imaging in pathology,'' \emph{Journal of pathology
  informatics}, vol.~2, 2011.

\bibitem{kumar2018deep}
M.~D. Kumar, M.~Babaie, and H.~R. Tizhoosh, ``Deep barcodes for fast retrieval
  of histopathology scans,'' in \emph{2018 International Joint Conference on
  Neural Networks (IJCNN)}.\hskip 1em plus 0.5em minus 0.4em\relax IEEE, 2018,
  pp. 1--8.

\bibitem{rajalocalization}
R.~S. Alomari, R.~Allen, B.~Sabata, and V.~Chaudhary, ``Localization of tissues
  in high-resolution digital anatomic pathology images,'' in \emph{Medical
  Imaging 2009: Computer-Aided Diagnosis}, vol. 7260.\hskip 1em plus 0.5em
  minus 0.4em\relax International Society for Optics and Photonics, 2009, p.
  726016.

\bibitem{khan2012ranpec}
A.~M. Khan, H.~El-Daly, and N.~Rajpoot, ``Ranpec: Random projections with
  ensemble clustering for segmentation of tumor areas in breast histology
  images,'' in \emph{Medical Image Understanding and Analysis}, 2012, pp.
  17--23.

\bibitem{he2012histology}
L.~He, L.~R. Long, S.~Antani, and G.~R. Thoma, ``Histology image analysis for
  carcinoma detection and grading,'' \emph{Computer methods and programs in
  biomedicine}, vol. 107, no.~3, pp. 538--556, 2012.

\bibitem{bandi2017comparison}
P.~B{\'a}ndi, R.~van~de Loo, M.~Intezar, D.~Geijs, F.~Ciompi, B.~van Ginneken,
  J.~van~der Laak, and G.~Litjens, ``Comparison of different methods for tissue
  segmentation in histopathological whole-slide images,'' in \emph{2017 IEEE
  14th International Symposium on Biomedical Imaging (ISBI 2017)}.\hskip 1em
  plus 0.5em minus 0.4em\relax IEEE, 2017, pp. 591--595.

\bibitem{babaie2019deep}
M.~Babaie and H.~R. Tizhoosh, ``Deep features for tissue-fold detection in
  histopathology images,'' \emph{arXiv preprint arXiv:1903.07011}, 2019.

\bibitem{kothari2013eliminating}
S.~Kothari, J.~H. Phan, and M.~D. Wang, ``Eliminating tissue-fold artifacts in
  histopathological whole-slide images for improved image-based prediction of
  cancer grade,'' \emph{Journal of pathology informatics}, vol.~4, 2013.

\bibitem{oswal2013entropy}
V.~Oswal, A.~Belle, R.~Diegelmann, and K.~Najarian, ``An entropy-based
  automated cell nuclei segmentation and quantification: application in
  analysis of wound healing process,'' \emph{Computational and mathematical
  methods in medicine}, vol. 2013, 2013.

\bibitem{Babaie_2017_CVPR_Workshops}
M.~Babaie, S.~Kalra, A.~Sriram, C.~Mitcheltree, S.~Zhu, A.~Khatami,
  S.~Rahnamayan, and H.~R. Tizhoosh, ``Classification and retrieval of digital
  pathology scans: A new dataset,'' in \emph{The IEEE Conference on Computer
  Vision and Pattern Recognition (CVPR) Workshops}, July 2017.

\bibitem{bentaieb2018predicting}
A.~BenTaieb and G.~Hamarneh, ``Predicting cancer with a recurrent visual
  attention model for histopathology images,'' in \emph{International
  Conference on Medical Image Computing and Computer-Assisted
  Intervention}.\hskip 1em plus 0.5em minus 0.4em\relax Springer, 2018, pp.
  129--137.

\bibitem{bug2015foreground}
D.~Bug, F.~Feuerhake, and D.~Merhof, ``Foreground extraction for
  histopathological whole slide imaging,'' in \emph{Bildverarbeitung f{\"u}r
  die Medizin 2015}.\hskip 1em plus 0.5em minus 0.4em\relax Springer, 2015, pp.
  419--424.

\bibitem{erfankhah2019heterogeneity}
H.~Erfankhah, M.~Yazdi, M.~Babaie, and H.~R. Tizhoosh, ``Heterogeneity-aware
  local binary patterns for retrieval of histopathology images,'' \emph{IEEE
  Access}, vol.~7, pp. 18\,354--18\,367, 2019.

\bibitem{otsu1982role}
T.~Otsu and M.~Yoshida, ``Role of initiator-transfer agent-terminator
  (iniferter) in radical polymerizations: Polymer design by organic disulfides
  as iniferters,'' \emph{Die Makromolekulare Chemie, Rapid Communications},
  vol.~3, no.~2, pp. 127--132, 1982.

\bibitem{mohit2017automated}
M.~Mohit, ``Automated histopathological analyses at scale,'' Ph.D.
  dissertation, Massachusetts Institute of Technology, 2017.

\bibitem{nguyen2011prostate}
K.~Nguyen, A.~K. Jain, and B.~Sabata, ``Prostate cancer detection: Fusion of
  cytological and textural features,'' \emph{Journal of pathology informatics},
  vol.~2, 2011.

\bibitem{fouad2017unsupervised}
S.~Fouad, D.~Randell, A.~Galton, H.~Mehanna, and G.~Landini, ``Unsupervised
  morphological segmentation of tissue compartments in histopathological
  images,'' \emph{PloS one}, vol.~12, no.~11, p. e0188717, 2017.

\bibitem{chen2019tissueloc}
P.~Chen and L.~Yang, ``tissueloc: Whole slide digital pathology image tissue
  localization.'' \emph{J. Open Source Software}, vol.~4, no.~33, p. 1148,
  2019.

\bibitem{ronneberger2015u}
O.~Ronneberger, P.~Fischer, and T.~Brox, ``U-net: Convolutional networks for
  biomedical image segmentation,'' in \emph{International Conference on Medical
  image computing and computer-assisted intervention}.\hskip 1em plus 0.5em
  minus 0.4em\relax Springer, 2015, pp. 234--241.

\bibitem{dong2017automatic}
H.~Dong, G.~Yang, F.~Liu, Y.~Mo, and Y.~Guo, ``Automatic brain tumor detection
  and segmentation using u-net based fully convolutional networks,'' in
  \emph{annual conference on medical image understanding and analysis}.\hskip
  1em plus 0.5em minus 0.4em\relax Springer, 2017, pp. 506--517.

\bibitem{bulten2018automated}
W.~Bulten, C.~A. Hulsbergen-van~de Kaa, J.~van~der Laak, G.~J. Litjens
  \emph{et~al.}, ``Automated segmentation of epithelial tissue in prostatectomy
  slides using deep learning,'' in \emph{Medical Imaging 2018: Digital
  Pathology}, vol. 10581.\hskip 1em plus 0.5em minus 0.4em\relax International
  Society for Optics and Photonics, 2018, p. 105810S.

\bibitem{naylor2018segmentation}
P.~Naylor, M.~La{\'e}, F.~Reyal, and T.~Walter, ``Segmentation of nuclei in
  histopathology images by deep regression of the distance map,'' \emph{IEEE
  transactions on medical imaging}, vol.~38, no.~2, pp. 448--459, 2018.

\bibitem{Yakubovskiy:2019}
P.~Yakubovskiy, ``Segmentation models,''
  \url{https://github.com/qubvel/segmentation_models}, 2019.

\bibitem{grayscal_regions}
S.~Jeyalaksshmi and S.~Prasanna, ``Measuring distinct regions of grayscale
  image using pixel values,'' \emph{International Journal of Engineering \&
  Technology}, vol.~7, no. 1.1, pp. 121--124, 2018.

\bibitem{belsare2012histopathological}
A.~Belsare and M.~Mushrif, ``Histopathological image analysis using image
  processing techniques: An overview,'' \emph{Signal \& Image Processing},
  vol.~3, no.~4, p.~23, 2012.

\bibitem{web_TCGA_dataset}
``The cancer genoum atlas (tcga) dataset,''
  \url{https://portal.gdc.cancer.gov/}, accessed: November 2019.

\bibitem{tomczak2015cancer}
K.~Tomczak, P.~Czerwi{\'n}ska, and M.~Wiznerowicz, ``The cancer genome atlas
  (tcga): an immeasurable source of knowledge,'' \emph{Contemporary oncology},
  vol.~19, no.~1A, p. A68, 2015.

\bibitem{howard2017mobilenets}
A.~G. Howard, M.~Zhu, B.~Chen, D.~Kalenichenko, W.~Wang, T.~Weyand,
  M.~Andreetto, and H.~Adam, ``Mobilenets: Efficient convolutional neural
  networks for mobile vision applications,'' \emph{arXiv preprint
  arXiv:1704.04861}, 2017.

\bibitem{simonyan2014very}
K.~Simonyan and A.~Zisserman, ``Very deep convolutional networks for
  large-scale image recognition,'' \emph{arXiv preprint arXiv:1409.1556}, 2014.

\bibitem{tan2019efficientnet}
M.~Tan and Q.~V. Le, ``Efficientnet: Rethinking model scaling for convolutional
  neural networks,'' \emph{arXiv preprint arXiv:1905.11946}, 2019.

\bibitem{he2016deep}
K.~He, X.~Zhang, S.~Ren, and J.~Sun, ``Deep residual learning for image
  recognition,'' in \emph{Proceedings of the IEEE conference on computer vision
  and pattern recognition}, 2016, pp. 770--778.

\bibitem{huang2017densely}
G.~Huang, Z.~Liu, L.~Van Der~Maaten, and K.~Q. Weinberger, ``Densely connected
  convolutional networks,'' in \emph{Proceedings of the IEEE conference on
  computer vision and pattern recognition}, 2017, pp. 4700--4708.

\bibitem{kingma2014adam}
D.~P. Kingma and J.~Ba, ``Adam: A method for stochastic optimization,''
  \emph{arXiv preprint arXiv:1412.6980}, 2014.

\end{thebibliography}
\end{document}